\documentclass{article}

\usepackage{PRIMEarxiv}

\usepackage[utf8]{inputenc} % allow utf-8 input
\usepackage[T1]{fontenc}    % use 8-bit T1 fonts
\usepackage{hyperref}       % hyperlinks
\usepackage{url}            % simple URL typesetting
\usepackage{booktabs}       % professional-quality tables
\usepackage{amsfonts}       % blackboard math symbols
\usepackage{nicefrac}       % compact symbols for 1/2, etc.
\usepackage{microtype}      % microtypography
\usepackage{lipsum}
\usepackage{graphicx}
\graphicspath{{media/}}     % organize your images and other figures under media/ folder
\usepackage{natbib}
  
\title{Uncovering Solar Wind Phenomena with iSAX, HDBSCAN, Human-in-the-loop and PSP Observations}

\author{%
 Valmir P. Moraes Filho \\
 Catholic University of America \\
 \texttt{moraesfilho@cua.edu} \\
 \And
 Daniela Martin \\
 University of Delaware \\
 \texttt{dmartinv@udel.edu}
 \And
 Jasmine R. Kobayashi\\
 Southwest Research Institute \\
 \texttt{jasmine.kobayashi@swri.org} \\
 \And
 Jinsu Hong \\
 Georgia State University \\
 \texttt{jhong36@gsu.edu} \\
 \And
 Connor O'Brien \\
 Boston University \\
 \texttt{connor.joseph.obrien@gmail.com} \\
 \And
 Evangelia Samara \\
 NASA Goddard Space Flight Center \\
 \texttt{evangelia.samara@nasa.gov} \\
 \And
 Joseph Gallego \\
 Drexel University \\
 \texttt{jg3959@drexel.edu} \\
}

\begin{document}
\maketitle

\begin{abstract}
The solar wind is a dynamic plasma outflow that shapes heliospheric conditions and drives space weather. Identifying its large-scale phenomena is crucial, yet the increasing volume of high-cadence Parker Solar Probe (PSP) observations poses challenges for scalable, interpretable analysis. We present a pipeline combining symbolic compression, density-based clustering, and human-in-the-loop validation. Applied to 2018–2024 PSP data, it efficiently processes over 150 GB of magnetic and plasma measurements, recovering known structures, detecting uncatalogued CMEs and transient events, and demonstrating robustness across multiple time scales. A key outcome is the systematic use of the magnetic deflection angle ($\theta_B$) as a unifying metric across solar wind phenomena. This framework provides a scalable, interpretable, expert-validated approach to solar wind analysis, producing expanded event catalogs and supporting improved space weather forecasting. The code and configuration files used in this study are publicly available to support reproducibility.
\end{abstract}

% keywords can be removed

\section{Introduction}
The solar wind is a continuous outflow of plasma from the Sun that shapes the heliosphere and drives space weather phenomena, which can affect satellites, astronauts, communications, and power grids \citep{Beedle_2022,Zank_2015,Kilpua_2017}. Key transient events within the solar wind include Coronal Mass Ejections (CMEs), High-Speed Streams (HSS), and Stream Interaction Regions (SIRs). CMEs are large-scale eruptions of magnetized plasma that can produce geomagnetic storms upon interacting with Earth’s magnetosphere \citep{Nieves-Chinchilla_2016,Kilpua_2015}. HSS arise from coronal holes and are often embedded within SIRs, regions where fast solar wind overtakes slower streams, leading to plasma compression, enhanced magnetic fields, and complex interface dynamics \citep{Jian_2006, Sanchez-Garcia_2024}. In addition, small-scale magnetic foldings known as switchbacks are ubiquitous in the near-Sun solar wind and exhibit rapid rotations of the magnetic field vector while maintaining near-constant plasma properties \citep{Kasper_2019}. Accurate detection and characterization of these diverse phenomena are crucial for space weather forecasting and for understanding solar wind dynamics.

The increasing volume of high-cadence in situ observations from missions like the Parker Solar Probe (PSP) poses significant challenges for traditional analysis methods \citep{raouafi2023parker}. Existing approaches include clustering and symbolic methods \citep{Roberts_2020,Camerra_2010}, deep learning-based detectors \citep{Nguyen_2019,Rudisser_2022,Nguyen_2025}, and physics- or threshold-based algorithms \citep{Hu_2018,Fargette_2021,Laker_2022,Raouafi_2023}. While these methods have demonstrated success in specific contexts, they are often limited to one or two types of events, computationally expensive, or difficult to interpret. Moreover, many techniques treat SIRs, CMEs, and switchbacks separately, missing opportunities to apply common diagnostics across phenomena.

To overcome these limitations, we use a pipeline combining iSAX symbolic compression \citep{Camerra_2010}, HDBSCAN clustering \citep{campello-etal-2013}, and human-in-the-loop validation \citep{kobayashi2025cipherscalabletimeseries}. PSP time series are segmented and encoded to preserve temporal patterns at multiple scales. Experts inspect representative windows, considering all relevant parameters, including the magnetic deflection angle $\theta_B$, which sharply rotates in switchbacks, varies gradually across SIRs, and rotates smoothly within CME ejecta. Labels from sampled windows are extrapolated to entire clusters, producing consistent, expert-validated catalogs. This approach enables scalable, interpretable identification of solar wind phenomena from 2018–2024 while highlighting $\theta_B$ as a unifying diagnostic across event types.

\section{Data}
PSP, launched in 2018, is the first spacecraft to approach the Sun as close as $10~R_\odot$ ($\approx 7$ million km), providing unprecedented in situ observations of the solar wind. In this study, we use high-cadence magnetic field measurements from the FIELDS instrument and ion plasma data from SWEAP (SPC and SPAN-I instruments) \citep{Bale_2016,Kasper_2016}. We focus on the following key parameters: magnetic field components ($B_r, B_t, B_n$), solar wind speed ($V_{SW}$), proton density ($N_P$), proton thermal speed ($V_P$).

The merged 2018–2024 dataset \citep{martin2025scalablemachinelearninganalysis}, exceeding 150 GB, is stored in compressed Zarr format \cite{zarr-github2025}. Linear interpolation fills small gaps, with flags preserved to indicate whether each value is observed or interpolated, enabling reliable anomaly detection. Distributed computing is supported via Dask \cite{dask}, facilitating scalable analysis of these solar wind structures across multiple time scales.

\section{Methodology}
\begin{figure}[hbtp]
    \centering
    \includegraphics[width=\linewidth]{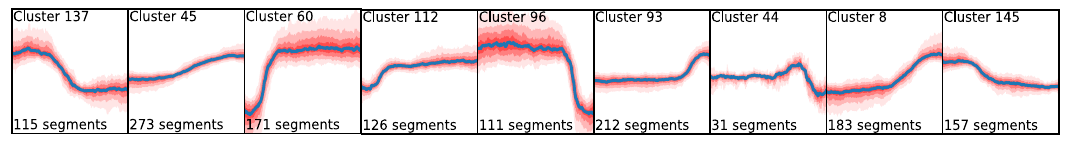}
    \caption{Representative summary of nine randomly selected clusters from one experiment, out of more than 150 total clusters. For each cluster, the mean time series is shown in blue, with the 5\%–95\% confidence intervals indicated in red shading. Cluster identifiers are indicated in the top-left corner, and the number of constituent windows (labeled as “segments” in the figure) is shown in the bottom-left corner. Across most clusters, variation around the mean is small, except for clusters 60 and 96, which exhibit slightly larger spread. Notably, the relationship between the mean and confidence intervals is clear, highlighting the characteristic temporal patterns captured within each cluster.}
    \label{fig:cluster-summary}
\end{figure}

Our pipeline consists of four steps, beginning with the preprocessing of the time series data. This step includes an optional detrending process, which removes large-scale trends, followed by smoothing to eliminate high-frequency noise \cite{monke2020optimal}. The second step involves compressing the time series using iSAX \citep{Camerra_2010}, which transforms Piecewise Aggregate Approximation (PAA) coefficients into symbolic “words” through statistically defined breakpoints. Each time series is divided into fixed-length windows, referred to as the \textit{chunk size}, and each window is further partitioned into smaller segments, or \textit{word size}, which define the temporal resolution of the symbolic encoding. This multi-resolution representation enables scalable indexing while preserving the essential temporal patterns. As the overall number of tree nodes becomes too large to be examined and categorized directly by an expert, clustering is performed on selected levels of the iSAX index using HDBSCAN \citep{campello-etal-2013}. HDBSCAN is a density-based algorithm that groups similar sequences while labeling low-density regions as noise. One hyperparameter, \textit{min\_cluster\_size}, defines the minimum number of sequences required to form a valid cluster, while another, \textit{min\_samples}, controls how strictly the algorithm classifies points in low-density regions as noise. To handle unassigned points, the pipeline can optionally re-cluster them under relaxed density constraints. Figure~\ref{fig:cluster-summary} illustrates representative clusters obtained from our pipeline. Nine randomly selected clusters from a given experiment are shown, highlighting the mean and confidence intervals of the constituent windows. 

 % This human-in-the-loop process enables the propagation of expert annotations across the entire cluster. 

For the final step, a domain expert evaluates representative windows from each cluster to validate physical consistency, resolve ambiguities, and assign meaningful labels. During this inspection, the expert considers all the corresponding PSP parameters for the same time intervals to ensure contextual coherence. The resulting labels are then propagated to all members of the cluster, yielding a robust expert-validated catalog of solar wind events.

In addition to the primary PSP measurements ($B_r$, $B_t$, $B_n$, $V_{SW}$, $N_P$, $V_P$), we include the derived parameters the total magnetic field ($|B|$), the magnetic deflection angle ($\theta_B$), and the proton temperature ($T_P$) to support the identification of shocks, SIRs, CMEs, and magnetic switchbacks. Among these, we hypothesize that $\theta_B$, defined from the three field components in the Radial–Tangential–Normal coordinate system ($B_R, B_T, B_N$) as the angle between the magnetic vector components \citep{Kasper_2019}, may provide particularly valuable information for characterizing solar wind events. As a compact measure of field orientation, $\theta_B$ captures both small variations indicative of radial alignment and large deviations corresponding to strong rotations, which could help differentiate abrupt switchbacks, smooth CME rotations, and steady SIR intervals. By applying this methodology, we structure the data into coherent groups that preserve the specificity of recurring solar wind patterns while keeping the labeling task scalable and interpretable. %Experts assign categories to representative windows, and these labels are extrapolated to all other cluster members, ensuring that labels reflect real solar wind phenomena, preserve physical consistency, resolve ambiguous cases, and produce robust, expert-validated catalogs of events.

\section{Experiments \label{sec:experiments}}
To capture solar wind events across multiple temporal scales, we analyzed PSP time series using windows ranging from 1 hour to 3 days. Longer windows (12–24 hours) were necessary to capture persistent structures such as SIRs, whereas shorter windows sufficed for transient phenomena, such as magnetic switchbacks. Each candidate interval was first cross-checked against existing CME catalogs \cite{richardson_cane_icmetable, helioforecast_icmecat} to exclude known events while also revealing previously uncatalogued ejecta, demonstrating that our pipeline can both reproduce and expand existing event inventories. Intervals not listed in catalogs were further examined by experts using field and plasma signatures, guided by established criteria for large-scale phenomena \cite{Jian_2006,Jian_2006a,Kilpua_2015,Nieves-Chinchilla_2016,Sanchez-Garcia_2024,Kasper_2019}.

\begin{figure}[!htbp]
    \centering
    \includegraphics[width=\textwidth]{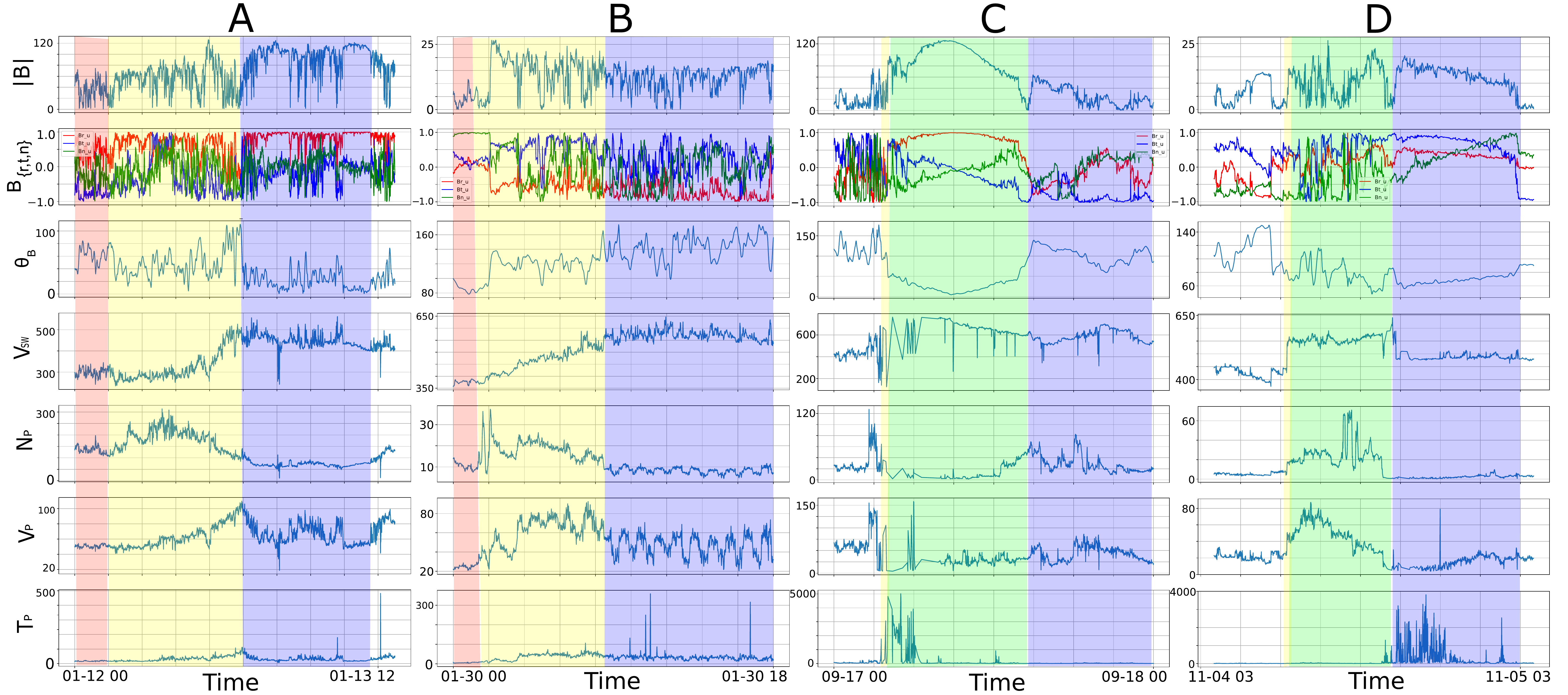}
    \caption{PSP observations of two SIRs (A and B: Jan 12–13, 2021, and Jan 30–31, 2022) and two CMEs (C and D: Sept 17–18 and Nov 3–4, 2023). For the SIRs, shaded regions indicate slow wind (red), compressed interaction region (yellow), and trailing HSS (blue). The compression region shows increased $|B|$, $N_P$, and $T_P$, while the HSS is faster, lower-density, and exhibits a steadier and more coherent $\theta_B$. The CME panels highlight contrasting structures: the September event features a forward shock (yellow), compressed sheath (green), and magnetic ejecta (purple) with smooth $\theta_B$ rotation, characteristic of a flux rope. The November event has a weak shock, modest compression, a turbulent sheath, and a magnetic ejecta with coherent field rotations, reduced speed, lower density, and suppressed temperature; $\theta_B$ shows moderate but coherent deflection across the ejecta.}
    \label{fig:PSP_events}
\end{figure}

Figure~\ref{fig:PSP_events} illustrates two representative SIR events identified using this approach. Panel~A shows the event of Jan 12–13, 2021, characterized by a leading compression region with enhanced $|B|$, $N_P$, and $T_P$, followed by a high-speed stream (HSS) exceeding 500 km/s with reduced density and more coherent magnetic field. Panel~B depicts the event of Jan 30-31, 2022, which exhibits a similar sequence, including a fast-stream plateau near 600 km/s. %In both cases, the magnetic field components ($B_r,B_t,B_n$) are more turbulent in the compression region and steadier within the HSS, consistent with interaction-region dynamics. 
These examples demonstrate that our methodology reproduces established SIR signatures \citep{Jian_2006, Sanchez-Garcia_2024, Richardson_2018} while enabling scalable, expert-validated identification of events across multiple temporal scales. Panel~C shows the Sept 16–17, 2023 CME, which begins with a forward shock marked by simultaneous jumps in $|B|$, $N_P$, $T_P$, and $V_{SW}$, followed by a hot, turbulent sheath, and a magnetic ejecta with enhanced $|B|$, smooth rotations in $B_{r,t,n}$, declining $V_{SW}$, and suppressed $T_P$, consistent with a classical magnetic cloud \cite{Nieves-Chinchilla_2016,Kilpua_2015}. Panel~D depicts the Nov 3–4, 2023 ICME, which lacks a pronounced shock and sheath, showing only a weak ejecta with moderately enhanced $|B|$, smooth magnetic field rotations, declining $V_{SW}$ from 600 to 450 km/s, reduced $N_P$, and low $T_P$ \cite{Jian_2006a}. Together, these CME cases illustrate how our pipeline captures contrasting CME morphologies, from strong events with classical shock-sheath-ejecta sequences to weaker, more subtle ejecta structures.

\begin{figure}[!h]
    \centering
    \includegraphics[width=\textwidth]{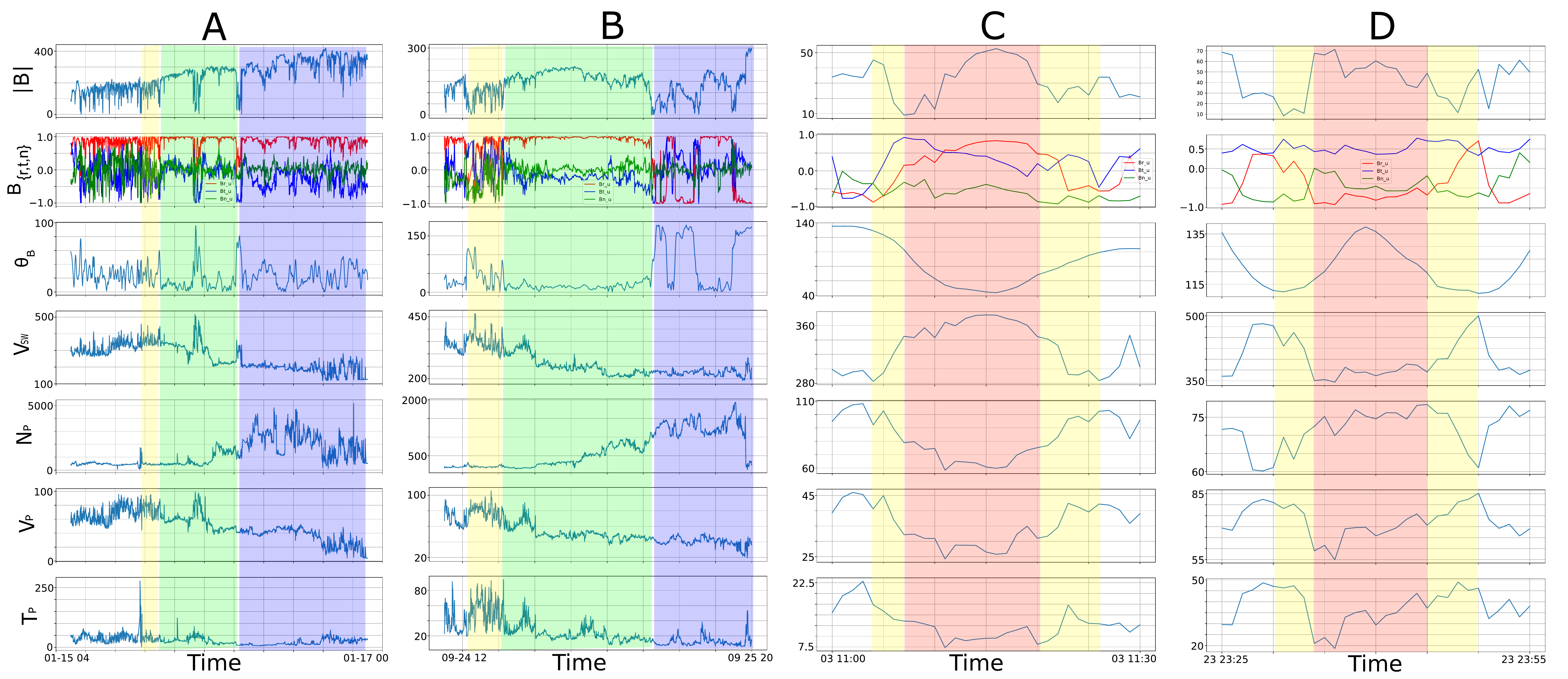}
    \caption{Parker Solar Probe observations of uncatalogued CMEs and magnetic switchbacks. 
Panels A-B: CMEs on Jan 15–16, 2021 and Sep 24–25, 2020, showing ejecta with enhanced $|B|$, smooth rotations in $B_{r,t,n}$, declining $V_{SW}$, and suppressed $T_P$, followed by trailing compressions with elevated $N_P$ and variable $\theta_B$. 
Panels C-D: switchbacks on Aug 3, 2021 and Dec 23, 2023, with shaded regions marking transition (yellow) and peak (red) deflections in $\theta_B$. Sharp rotations occur in both cases ($\sim$90° in August, 60°–80° in December), while plasma parameters remain stable, confirming their Alfvénic nature \cite{Kasper_2019}.}
    \label{fig:PSP_events2}
\end{figure}

Figure~\ref{fig:PSP_events2} illustrates two uncatalogued CME cases with distinct morphologies. Panel~A shows the event of Jan 15–16, 2021, in which PSP observed a clear magnetic cloud with enhanced $|B|$, smooth rotations in $B_{r,t,n}$, declining $V_{SW}$, and suppressed $T_P$, preceded by a hot, dense sheath—well-known CME signatures \cite{Burlaga_1981, Nieves-Chinchilla_2016, Kilpua_2015}. Panel~B presents the event of Sep 24–25, 2020, which, in contrast, exhibited a weaker ejecta with moderate $|B|$ enhancement, coherent field rotations, and declining $V_{SW}$, accompanied by elevated $N_P$ and disturbed plasma in the trailing region, consistent with a compressed sheath \cite{Jian_2006a, Kilpua_2017}. These uncatalogued events confirm the pipeline’s ability to detect CMEs with diverse ejecta and sheath properties, in line with established CME morphology \cite{Richardson_2010}. Finally, switchbacks were captured, confirming the sensitivity of the method to short-lived structures, see Figure~\ref{fig:PSP_events2} (Panel C and D). On Aug 3, 2021, PSP observed close to $90^\circ$ rotation in $\theta_B$, nearly reversing the radial field. On Dec 23, 2023, $\theta_B$ rotated by $60^\circ$ to $80^\circ$, producing a strong but partial deflection. In both cases, plasma properties ($V_{SW}, N_P, V_P, T_P$) remained stable, consistent with switchbacks as Alfvénic magnetic folds rather than compressive structures \cite{Kasper_2019}.

Across all event types, $\theta_B$ proved especially valuable. In switchbacks, sharp rotations of $60^\circ$ to $90^\circ$ are their defining Alfvénic signature \cite{Kasper_2019}. In SIRs, $\theta_B$ shows gradual variations across the stream interface \cite{Jian_2006}, while in CMEs it captures the smooth, large-scale rotations within magnetic ejecta \cite{Jian_2006a,Sanchez-Garcia_2024}. Although these applications have typically been treated separately in the literature, our results show that $\theta_B$ provides a consistent diagnostic across different solar wind phenomena. To our knowledge, this broader perspective suggests that $\theta_B$ can serve as a common metric for identifying and distinguishing solar wind transients.

\textbf{Hardware.} All experiments were conducted on Google Cloud Platform using four c2-standard-8 VM instances (8 vCPUs, 64 GB RAM) with 4 TB of SSD storage.

\section{Conclusion}
We presented a scalable and interpretable pipeline for identifying large-scale solar wind phenomena in PSP data. By combining iSAX symbolic compression, HDBSCAN clustering, human expert validation, and the magnetic deflection angle, $\theta_B$, the method efficiently analyzes dense, high-cadence datasets. Applied to PSP observations from 2018–2024, it successfully recovers SIRs, CMEs, and switchbacks, while also detecting previously uncatalogued events, supporting the creation of an expert-validated catalog. A central finding is the systematic use of $\theta_B$ as a unifying diagnostic across solar wind regimes: switchbacks exhibit sharp excursions, CME ejecta show smooth rotations, and SIRs display gradual variations. This demonstrates that $\theta_B$ provides a consistent, versatile metric for distinguishing transients that are often studied in isolation.

\section*{Broader Impact}
The framework reduces computational cost while preserving interpretability, making it well-suited for heliophysics missions and large-scale time series datasets. We plan to produce expert-validated event catalogs that advance our understanding of the solar wind structure and support future space weather efforts. The full codebase supporting this work is available at \url{https://github.com/spaceml-org/CIPHER}.

\section*{Acknowledgements}
This work is a research product of Heliolab (heliolab.ai), an initiative of the Frontier Development Lab (FDL.ai). FDL is a public–private partnership between NASA, Trillium Technologies (trillium.tech), and commercial AI partners including Google Cloud and NVIDIA. Heliolab was designed, delivered, and managed by Trillium Technologies Inc., a research and development company focused on intelligent systems and collaborative communities for Heliophysics, planetary stewardship and space exploration. We gratefully acknowledge Google Cloud for extensive computational resources enabled through VMware. This material is based upon work supported by NASA under award No. 80GSFC23CA040. Any opinions, findings, and conclusions or recommendations expressed are those of the author(s) and do not necessarily reflect the views of the National Aeronautics and Space Administration.

% Deep learning classifiers have achieved great success in pattern recognition tasks in numerous fields \citet{LeCun2015}. Relationships that can be checked: velocity-density and velocity-temperature for all heliocentric distances of PSP but also with 0.1 au radial step. Also, distributions of velocity, density, temperature, magnetic field with a radial step of 0.1 au.

%Bibliography
\bibliographystyle{unsrt}  
\bibliography{references}

\end{document}